\documentclass[%
reprint,
superscriptaddress,
showpacs,preprintnumbers,
nofootinbib,
amsmath,amssymb,
aps,
prl
showkeys
]{revtex4-1}

\usepackage{graphicx}
\usepackage{dcolumn}
\usepackage{bm}

\usepackage{amssymb}
\usepackage{amsmath}
\usepackage{amsfonts}
\usepackage{xspace}
\usepackage{bm,bbm}
\usepackage{relsize}
\usepackage{siunitx}
\usepackage[usenames,dvipsnames]{color}
\usepackage{float}

\renewcommand{\vec}[1]{\ensuremath{\bm{#1}}}

\newcommand{\ie}{\emph{i.e.}\xspace~}

\newcommand{\eg}{\emph{e.g.}\xspace~}

\newcommand{\identity}{\ensuremath{\mathlarger{\mathbbm{1}}}}
\renewcommand{\eqref}[1]{Eq.~\ref{eq:#1}}

\newcommand{\eq}{\ensuremath{\,{=}\,}}
\newcommand{\defas}{\ensuremath{\,{:=}\,}}

\usepackage{mathtools} 
\DeclarePairedDelimiter\bra{\langle}{\rvert}
\DeclarePairedDelimiter\ket{\lvert}{\rangle}
\DeclarePairedDelimiterX\braket[2]{\langle}{\rangle}{#1 \delimsize\vert #2}

\begin{document}

\title{
Gapless Unidirectional Photonic Transport Using All-Dielectric Kagome Lattices
} 

\author{Stephan Wong}
\email[Email: ]{wongs16@cardiff.ac.uk}
\affiliation{School of Physics and Astronomy, Cardiff University, Cardiff CF24 3AA, UK}
\affiliation{The Blackett Laboratory, Imperial College London, London SW7 2AZ, UK}
\author{Matthias Saba}
\affiliation{The Blackett Laboratory, Imperial College London, London SW7 2AZ, UK}
\affiliation{Adolphe Merkle Institute, University of Fribourg, 1700 Fribourg, Switzerland}
\author{Ortwin Hess}
\affiliation{The Blackett Laboratory, Imperial College London, London SW7 2AZ, UK}
\affiliation{School of Physics, CRANN, \& AMBER, Trinity College Dublin, Dublin 2, Ireland}
\author{Sang Soon Oh}
\email[Email: ]{ohs2@cardiff.ac.uk}
\affiliation{School of Physics and Astronomy, Cardiff University, Cardiff CF24 3AA, UK}

\date{\today}

\begin{abstract}
Photonic topological insulators are a promising photonic platform due to the possibility of unidirectional edge states with insensitivity to bending, fabrication imperfections or environmental fluctuation. 
Here we demonstrate highly efficient unidirectional photonic edge mode propagation facilitated by an optical analogue of the quantum valley Hall effect. 
With an all-dielectric kagome lattice design, we demonstrate broadband suppressed reflection in the presence of sharp corners and further show negligible vertical losses in a semiconductor-based device at telecommunication wavelengths.
\end{abstract}

\keywords{
}

\maketitle

When propagating in a (structured) material or waveguide, not all of the light travels in this initial direction but parts of it experience such back-reflection due to bending, fabrication defects or environmental variations. For most applications back-propagation should be avoided and it is thus not surprising that the unique properties of photonic topological insulators (PTIs)~\cite{Lu2016, Ozawa2019} have attracted widespread attention due to their promise to prohibit back-reflections. The basis of such back-scattering-free one-way waveguides lies at the interface of two topologically inequivalent photonic crystals (PhCs) which exhibit topological edge modes that -- guaranteed by the bulk-boundary correspondence~\cite{Hatsugai1993a} -- propagate only in one direction and are at the same time robust against perturbations. 
Not surprisingly, a plethora of possible topologically non-trivial photonic designs has been put forward, involving 
non-reciprocal systems~\cite{Haldane2008}, 
complex metamaterials~\cite{Khanikaev2012}, 
the Floquet topological insulator principle~\cite{Rechtsman2013},
and an artificial magnetic gauge~\cite{Hafezi2011, Fang2012}. 
However, the aforementioned PTIs need strong magnetic fields, are complicated to fabricate, and/or are difficult if not impossible to scale to optical frequencies.  

As an alternative, a deformed honeycomb-based topological PhC~\cite{Wu2015} which emulates the quantum spin Hall effect (QSHE)~\cite{Kane2005, Wu2015, Chen2018_QSHE, Yang2018} has recently gained interest, not least due to its simple fabrication as compared to other PTIs.  
Nevertheless, while 2D hexagonal symmetries (such as the honeycomb-based topological PhC) generally lead to Dirac cones at the K and K$'$ points of the Brillouin zone (BZ), 
and with a geometrical perturbation it is possible to lift the point-like degeneracies in order to obtain a non-trivial topological and complete photonic band gap~\cite{Saba2019} (which leads to topological protection defined within the parameter space of a certain type of a deterministic geometrical perturbation that differs from the traditional Hatsugai sense~\cite{Hatsugai1993a}), there is an inherent problem. 
The pseudo-time-reversal anti-unitary operator $\mathcal{T}^2 \eq -\identity$, introduced to have well-defined orthogonal spin up/down channels, is constructed on the basis of the six-fold rotation ($C_6$) operator of the crystal. However, the $C_{6}$ symmetry of the crystal is broken in any finite, truncated, configuration and the spin up and spin down channels couple to each other.
Consequently, while edge modes are guaranteed at the interface between the two topologically distinct deformed honeycomb PhCs, for most frequencies within the band gap, there is an anti-crossing in their dispersion and they eventually do suffer from intrinsic back-reflection.
Yet, starting from a $C_{6v}$ symmetry with symmetry protected Dirac cones, it can be shown that there are two routes towards breaking the symmetry to open a topological band gap without breaking optical reciprocity~\cite{Saba2019}: one related to the aforementioned QSHE~\cite{Kane2005}, and the other to the quantum valley Hall effect (QVHE)~\cite{Kim2014}.
The QVHE has been widely studied in 
photonic
and mechanical
systems with staggered honeycomb~\cite{Chen2017, Dong2017, Bleu2017, Chen2018, Gao2017, He2019, Pal2017} lattice, triangular rods~\cite{Ma2016, Shalaev2018, Kang2018} or multi-pod~\cite{Gao2018, Liu2019, Qian2018} structure.

Here, we introduce an all-dielectric PTI based on a kagome lattice~\cite{Syozi1951} that naturally lends itself to QVHE symmetry breaking~\cite{Saba2019} while being composed of monodisperse rods of a single dielectric material. 
Compared to the QVHE designs of triangular-like holes/rods array and staggered honeycomb structures, 
the proposed perturbed kagome lattice requires a single monodisperse type of circular holes/rods and is thus easier fabricated.
It is this simplicity of the kagome-based design in terms of fabrication and its unidirectional edge mode transport which makes it an ideal candidate for practical applications at near-infrared and visible wavelengths. 
Here, we present and model the predicted behaviour for an on-chip platform that can be readily fabricated with state-of-the-art semiconductor growth techniques~\cite{Kim2016a}.

\begin{figure}[t]
\centering
\includegraphics[width=\columnwidth]{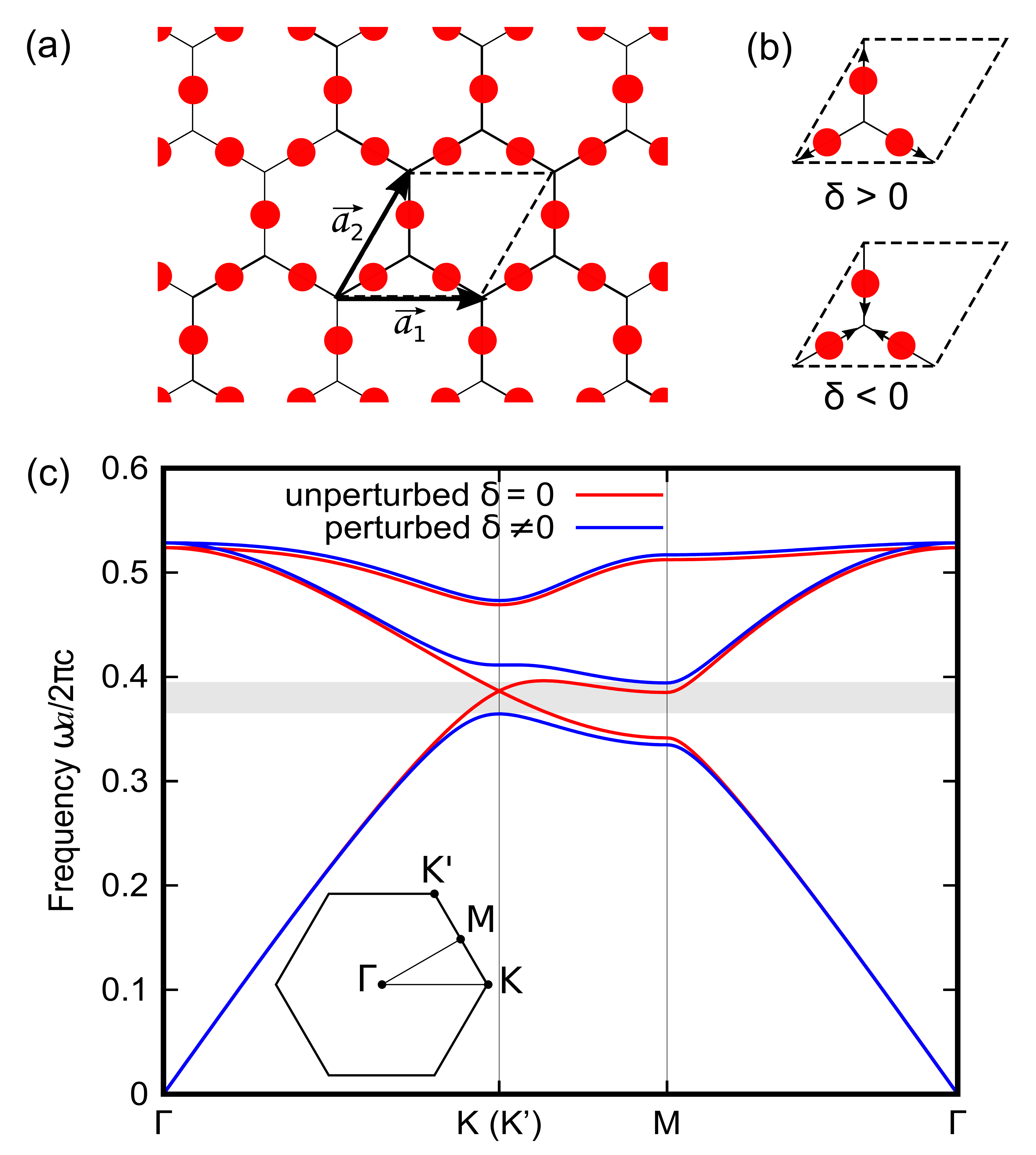}
\caption{
(a) Schematic of a kagome lattice. The solid line is a guide to the eye for the hexagonal symmetry. 
(b) Perturbations are considered by putting the rods further away (top) or closer (down) to each other, represented by the arrows.  
(c) Band structure of the kagome lattice for the unperturbed (solid red line) and perturbed (solid blue line) cases. The inset shows the first Brillouin zone.
}
\label{fig:kagome}
\end{figure}

The kagome lattice, named after a traditional Japanese basketweave pattern~\cite{Syozi1951}, has lattice sites at the midpoints of the edges in the regular hexagonal wallpaper tiling $\{6,3\}$, as illustrated in Fig.~\ref{fig:kagome}(a).
The unit cell is here composed of three rods and the perturbation to lift the degeneracy can be introduced such that these rods get closer (negative perturbation $\delta \, {<} \, 0$) or further away (positive perturbation $\delta \, {>} \, 0$) from their shared corner of the hexagons (Fig.~\ref{fig:kagome}(b)): $\vec{r} \mapsto \vec{r'} \eq (1 \pm \delta) \vec{r}$ where $\vec{r}$ is a vector taken from the corner of a hexagon to the adjacent rod. 

Importantly, the perturbation opens a band gap by lifting the linear degeneracy at the K and K$'$ points which is symmetry-induced in the unperturbed case. 
This generic behaviour manifests itself in the photonic bandstructure for $E_z$ out-of-plane transverse magnetic (TM) polarization (Fig.~\ref{fig:kagome}(c)) obtained with the open-source software \textit{MIT Photonic Bands} (MPB)~\cite{Johnson2001} for the unperturbed $\delta \eq 0$ and perturbed $\delta \eq {\pm} \, 0.15$ case. 
The model kagome photonic crystal, with lattice constant $a$, is composed of rods with diameter $d \eq 0.2a$ and permittivity $\varepsilon \eq 12$ within a vacuum ($\varepsilon \eq 1$) background. 
In contrast to the perturbed honeycomb PhC~\cite{Wu2015}, the translation symmetry is conserved after the perturbation, \ie the primitive lattice vectors (albeit not unique) are themselves preserved. 
As a consequence, the reciprocal lattice vectors remain the same and therefore the K and K$'$ points remain distinct and do not map to the $\Gamma$ point as is the case for the perturbed honeycomb lattice~\cite{Wu2015}.

To describe the band gap opening more rigorously, we adopt a perturbation-based group theoretical approach detailed in Ref.~\cite{Saba2017, Saba2019} to the hexagonal wallpaper group \textit{p6mm (17)}. Then, an effective Hamiltonian is derived for a (small) geometrical perturbation, $\tilde{\delta}$, and close to K, $\delta \vec{k} \eq (\delta k_x, \delta k_y)$, in the canonical basis of the induced K irreducible representation (irrep) from the 2D little group irrep of $p6mm$~\cite{Bradley}:
\begin{equation}
\mathcal{H}_\text{kagome} = \delta k_x \gamma_1 - \delta k_y \gamma_2 + \tilde{\delta} \gamma_5 = 
\left(
\begin{array}{cc}
\mathcal{W}_\text{K} & 0 \\ 
0 & \mathcal{W}_{\text{K}'}
\end{array} 
\right)
\label{eq:H_kagome}
\end{equation}
where $\gamma_1 \defas \sigma_3 \, {\otimes} \, \sigma_3$, $\gamma_2 \defas \sigma_3 \, {\otimes} \, \sigma_1$, $\gamma_3 \defas \sigma_1 \, {\otimes} \, \identity_2$, $\gamma_4 \defas \sigma_2 \, {\otimes} \, \identity_2$, $\gamma_5 \defas \sigma_3 \, {\otimes} \, \sigma_2$ are matrices satisfying a Clifford algebra generated by the vector field spanned by the $\gamma_i$, with the associated anti-commutation relation $\{\gamma_i, \gamma_j\} \eq 2 \delta_{ij} \identity_4$. $\mathcal{W}_{\text{K}/\text{K}'} \eq \pm \vec{h} \cdot \vec{\sigma}$ represent Weyl Hamiltonians of opposite chirality in the vicinity of K/K$'$ with $\vec{h} \eq (-\delta k_y, \tilde{\delta}, \delta k_x)$ and $\vec{\sigma} \eq (\sigma_1, \sigma_2, \sigma_3)$ the Pauli matrix vector. 
In Eq.~\ref{eq:H_kagome}, $\tilde{\delta}$ models the geometrical perturbation and therefore is proportional to the displacement of the rods $\delta$ away from the one of the unperturbed lattice.
The eigenvalues $E$ of the Weyl Hamiltonian at the K/K$'$ point are $E \eq {\pm} \, \sqrt{\delta k_x^2 + \delta k_y^2 + \tilde{\delta}^2}$ confirming that we obtain a linear degeneracy (Dirac point) in the 2D BZ for $\tilde{\delta} \eq 0$, \ie for the unperturbed kagome lattice, and have a band gap for $\tilde{\delta} \, {\neq} \, 0$.

Additionally, from Eq.~\ref{eq:H_kagome} it is clear that the effective Hamiltonian $\mathcal{H}_\text{kagome}$ is necessarily block-diagonalized~\cite{Bradley} in its canonical basis as in the case with the QSHE Kane-Mele Hamiltonian~\cite{Kane2005}, because translation symmetry is not broken with the perturbation. It does not mix the K/K$'$ irrep of the invariant translation group. 
The K/K$'$ points thus play the role of two orthogonal pseudo-spin channels, known as the valley degree of freedom, with the generating unitary pseudo-time-reversal operator $\mathcal{\tilde{T}} \eq \sigma_3 \, {\otimes} \, \identity_2$ with $[ \mathcal{\tilde{T}}, \mathcal{H}_\text{kagome}] \eq 0$, invariant under rotational symmetry breaking. 

\begin{figure}[t]
    \centering
    \includegraphics[width=\columnwidth]{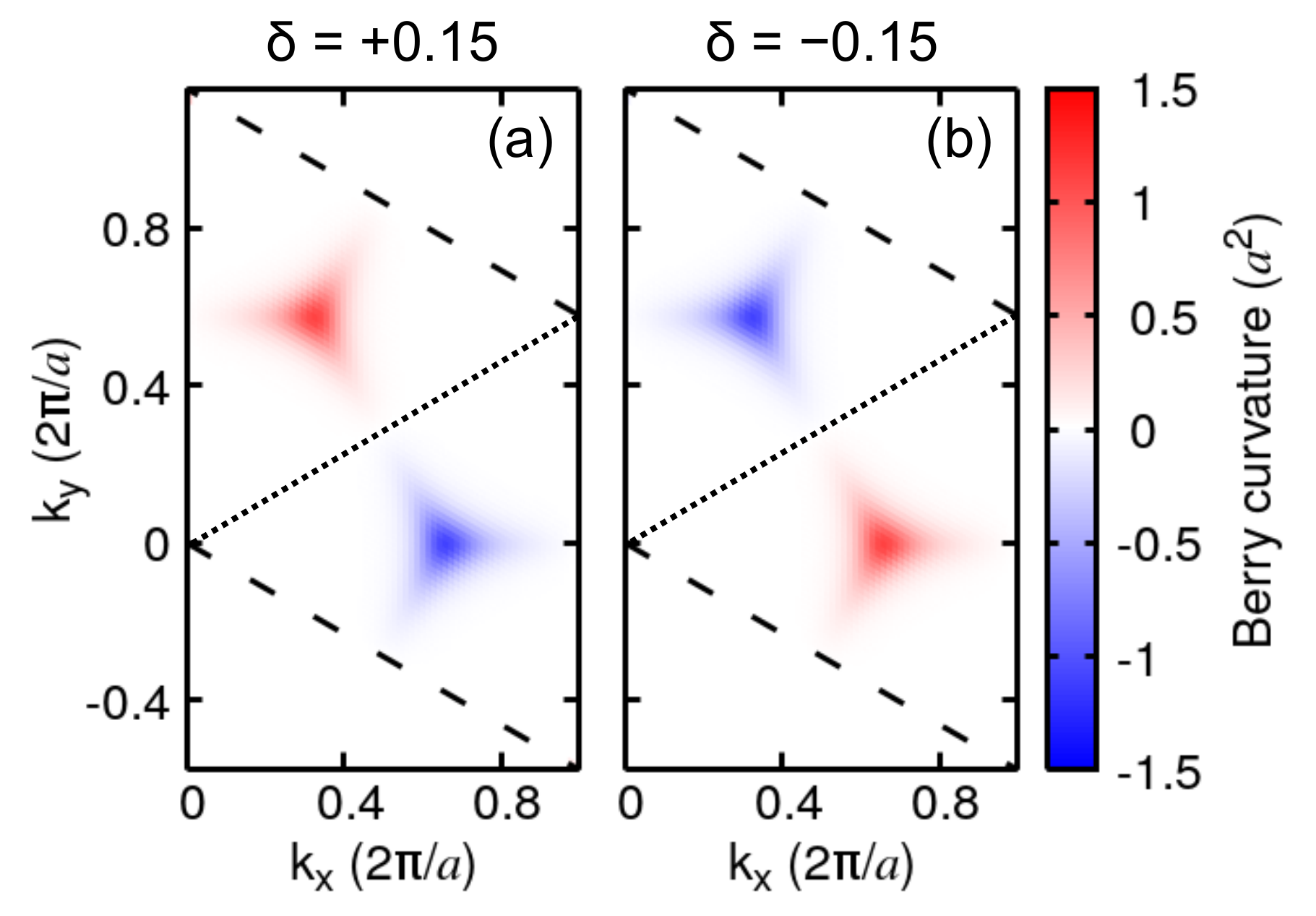}
    \caption{
        Calculated Berry curvatures in the reciprocal primitive unit cell, using a Wilson loop approach for (a) positively and (b) negatively perturbed kagome lattices. The dashed line represents the limit of the reciprocal primitive unit cell and the dotted black line defines the boundary for the regions of integration $S_{\text{K}/\text{K}'}$ in Eq.~\ref{eq:valley_chern_nb}.
    }
    \label{fig:valley_chern_nb}
\end{figure}

To show how the proposed kagome design emulates the QVHE emerging from the non-trivial Dirac points in the 2D BZ, we use the valley Chern number defined at the K and K$'$ points~\cite{Chen2017, Dong2017, Bleu2017, Chen2018, Shalaev2018, Qian2018, Kang2018, Liu2019, Gao2018, Gao2017, He2019, Ma2016}: 

\begin{equation}
C_{\text{K}/\text{K}'} = \frac{1}{2\pi} \int_{S_{\text{K}/\text{K}'}} \vec{\mathcal{F}(k)} \; d^2\vec{k}
.
    \label{eq:valley_chern_nb}
\end{equation}
The integration is performed over the two valley domains $S_{\text{K}/\text{K}'}$ defined in Fig.~\ref{fig:valley_chern_nb} as the two triangles which together form the dual of the standard hexagonal BZ tessellation of reciprocal space. The integrand is the Berry curvature~\cite{Berry1984} $\vec{\mathcal{F}(k)} \eq \nabla \times i \braket{\vec{u_k}}{\nabla_{\vec{k}}\vec{u_k}}$. Note that $C_{\text{K}/\text{K}'}$ is not a quantized topological invariant because $S_{\text{K}/\text{K}'}$ is not a closed surface, so that the Chern theorem does not hold.
Figure~\ref{fig:valley_chern_nb} shows the Berry curvatures calculated for a positively ($\delta \eq {+} \, 0.15$) and negatively ($\delta \eq {-} \, 0.15$) deformed kagome lattice, for $\vec{k}$ points in the reciprocal primitive unit cell. They are directly calculated from the numerical field profile using a Wilson loop approach to numerically avoid random gauge dependence (see Supplemental Material~\cite{Supp} for the details of the calculation). The valley Chern number obtained for positive (negative) perturbation by integrating around the K/K$'$ are then $C_{\text{K}/\text{K}'} \eq {\pm} \, 0.18$ ($C_{\text{K}/\text{K}'} \eq {\mp} \, 0.18$). 
It can be shown that the valley Chern number depends on the perturbation strength, and explains why we can only get a valley Chern number $C_{\text{K}/\text{K}'} \eq {\pm} \, 1/2$ for infinitesimal perturbation, as reported before~\cite{Chen2017, Dong2017, Bleu2017, Chen2018, Shalaev2018, Qian2018, Kang2018, Liu2019, Gao2018, Gao2017, He2019, Ma2016} (see Supplemental Material\cite{Supp} for more details on the valley Chern number).

\begin{figure}[t]
\centering
\includegraphics[width=\columnwidth]{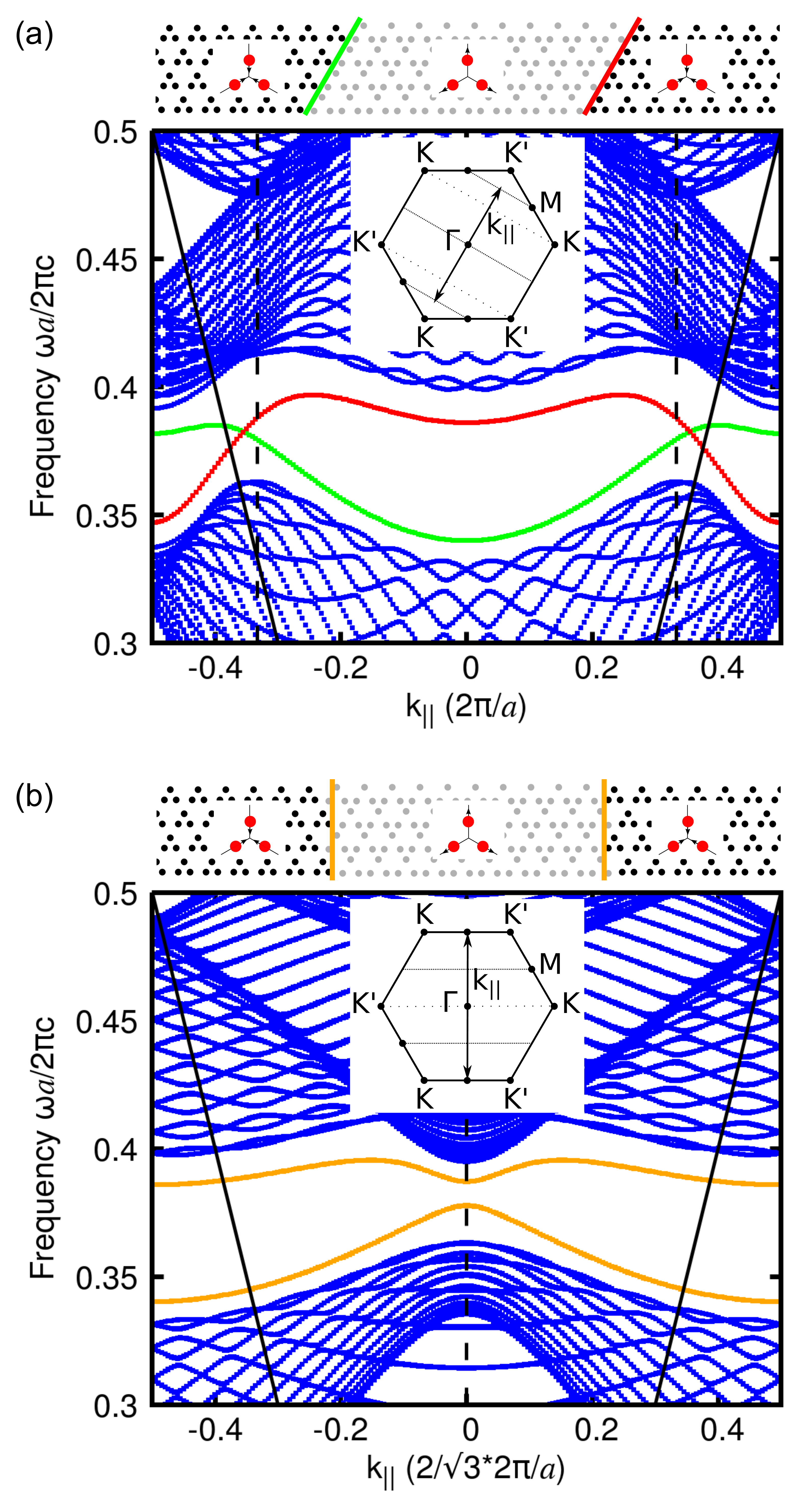}
\caption{
Supercell band structure at the interface between two topologically distinct kagome photonic crystals along different inclination directions: (a) $\Gamma {-} \text{K}$, (b) $\Gamma {-} \text{M}$.  The front and back interfaces are represented by solid lines with different colours on both the sketch and the band structure in (a). 
In (b), the interfaces and the edge mode dispersions are represented by the same colour because the two interfaces are equivalent.
The insets show the $k_\parallel$ sweep direction. The solid black line represents the light lines. 
The vertical dashed lines mark the position of K/K$'$.
The parameters are the same as in Fig.~\ref{fig:kagome} and with perturbation strength $\delta \eq {\pm} \, 0.15$.
}
\label{fig:supercell_band}
\end{figure}

Although the valley Chern number is not a topological invariant, a strong bulk-boundary correspondence similar to Ref.~\cite{Hatsugai1993a} exists in the extended parameter space $(-\delta k_y,\tilde{\delta}, \delta k_x)$ where Weyl charges of opposite chirality lead to guaranteed edge modes in the K/K$'$ valleys, respectively~\cite{Saba2019}. These Weyl charges can be correlated one-to-one to the sign of the valley Chern numbers, which can thus be interpreted as a topological integer with associated bulk-boundary correspondence.
In contrast to Ref.~\cite{Hatsugai1993a}, however, the existence of the strong correspondence in the extended parameter space only fixes crystal termination and is only valid for inclinations for which the K and K$'$ point are not projected to the same point in the edge BZ (cf. BZ insets in Fig.~\ref{fig:supercell_band}). The bulk-boundary correspondence thus reduces to the weaker form which rigorously valid only for specific well-defined boundaries.
From the expression of the effective Hamiltonian $\mathcal{H}_\text{kagome}$, it is evident that a non-trivial Weyl charge is located at the K and K$'$ points which has opposite signs because $\mathcal{W}_\text{K} \eq {-} \, \mathcal{W}_{\text{K}'}$ and which has opposite signs for opposite perturbation strength.
As a consequence, and from Fig.~\ref{fig:valley_chern_nb}, 
starting with an unperturbed lattice and then perturbing positively on one side of a chosen interface and negatively on the other could potentially lead to topological edge modes at most frequencies.

Figure~\ref{fig:supercell_band} shows the corresponding supercell band structure for $\delta \eq {\pm} \, 0.15$ and for different interface inclinations. 
The solid blue lines correspond to the bulk modes and the solid coloured lines correspond to topological edge modes inside the bulk band gap. 
This figure shows that an anti-crossing arises when the K/K$'$ points are projected onto the same $k_\parallel$ points ($\Gamma {-} \text{M}$ inclination, Fig.~\ref{fig:supercell_band}(b)) while a crossing arises for $\Gamma {-} \text{K}$ inclination, (Fig.~\ref{fig:supercell_band}(a)). 
In the $\Gamma {-} \text{K}$ inclination case (Fig.~\ref{fig:supercell_band}(a)), each interface supports two counter-propagating edge modes corresponding to the well-defined pseudo-spin up (down) edge modes of the K (K$'$) valleys. These result from the opposite Weyl charges at the K (or K$'$) point for two sides of the interface~\cite{Saba2019, Hatsugai1993a}. %
In the $\Gamma {-} \text{M}$ inclination case (Fig.~\ref{fig:supercell_band}(b)), however, the pseudo-spin separation breaks down and the edge modes suffer from back-scattering, similarly to the honeycomb PhC~\cite{Wu2015}. 

\begin{figure}[t]
\centering
\includegraphics[width=\columnwidth]{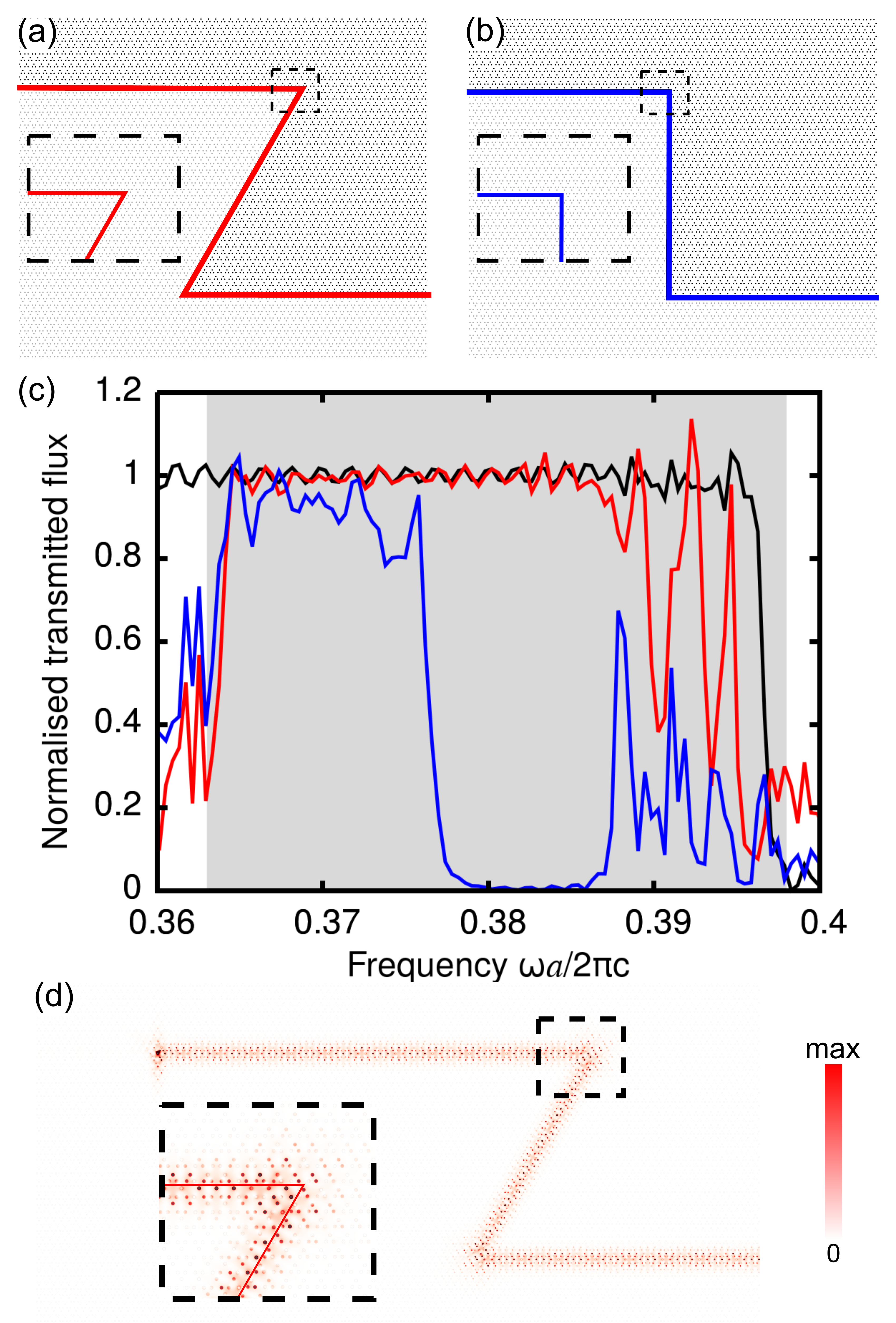}
\caption{
(a)-(b) Waveguides where the bending has a: (a) $Z$-shape or (b) staircase-shape where the bright (dark) rods are the positively (negatively) perturbed kagome lattice. Insets are zoomed in of the interface at the dashed square region.
(c) Transmission spectra for a waveguide without any bending (solid black line), $Z$-shaped (solid red line) or staircase-shaped (solid blue line) bending. The shaded area is a guide for the band gap frequency range.  
The parameters are kept the same as in Fig.~\ref{fig:supercell_band}. 
(d) Power of the edge modes excited at the $Z$-shaped bending structure.
}
\label{fig:transmission}
\end{figure}

We here demonstrate broadband back-scattering-immunity numerically in Fig.~\ref{fig:transmission} for finite perturbations, going beyond what has been shown rigorously for infinitesimal perturbation strengths with our generic theory based on symmetry only.
The unidirectional propagation with negligible inter-valley coupling is demonstrated by studying the transmission through waveguides with bendings of different inclinations.
Figure~\ref{fig:transmission}(a),(b) shows two examples of waveguides oriented in a $\Gamma {-} \text{K}$ inclination direction in which a bending is introduced such that the projected wavevectors of K and K$'$ onto the $k_\parallel$ line are distinct in a $Z$-shaped design (Fig.~\ref{fig:transmission}(a), inset Fig.~\ref{fig:supercell_band}(a)), or fall onto the same point along the vertical interface in a staircase-shaped design (Fig.~\ref{fig:transmission}(b), inset Fig.~\ref{fig:supercell_band}(b)).
In the case of the $Z$-shaped waveguide we focus on the red interface highlighted in Fig.~\ref{fig:supercell_band}(a) (which is not equivalent to the green interface). 
Figure~\ref{fig:transmission}(c) shows the normalised transmitted flux at the end of the waveguides, obtained using \textit{MIT Electromagnetic Equation Propagation (MEEP)}~\cite{Oskooi2010}.
At high frequencies, around the upper edge of the photonic bulk band gap, large oscillations are observed due to the presence of additional modes that  couple with the two valley channels (see Fig.~\ref{fig:supercell_band} and Supplemental Material~\cite{Supp} for more information on the modal transmissions).
The transmission for the $Z$-shaped waveguide (solid red line) is of the same order of magnitude as for a straight waveguide with no bending (solid black line) for most of the frequencies inside the band gap (represented by the grey shaded region). 
In contrast, the staircase-shaped  bending (solid blue line) leads to substantial back-reflection resulting in lower transmission because of the non-negligible inter-valley couplings on the $\Gamma {-} \text{M}$ inclination interface. This $\Gamma {-} \text{M}$ inclination bending additionally introduces a band gap in the dispersion where the edge mode cannot propagate.
Broadband robust transmission is therefore achieved for the $\Gamma {-} \text{K}$ inclination and in practice the orthogonality of two pseudo-spin channels at the wavevectors away from K/K$'$ is as good as at K/K$'$.
Figure~\ref{fig:transmission}(d) shows the power profile for a mode propagating along the interface with bendings at frequency $\omega \eq 0.37 \, (2\pi c/a)$ with negligible back-reflection. 
Here, the unidirectional edge mode has been excited using a rotating magnetic point dipole source~\cite{Lodahl2017, Wu2015} and the spatial position of the source has been determined using a chirality map~\cite{Oh2018} (see Supplemental Material~\cite{Supp} for more details on the chirality map).

\begin{figure}
\centering
\includegraphics[width=\columnwidth]{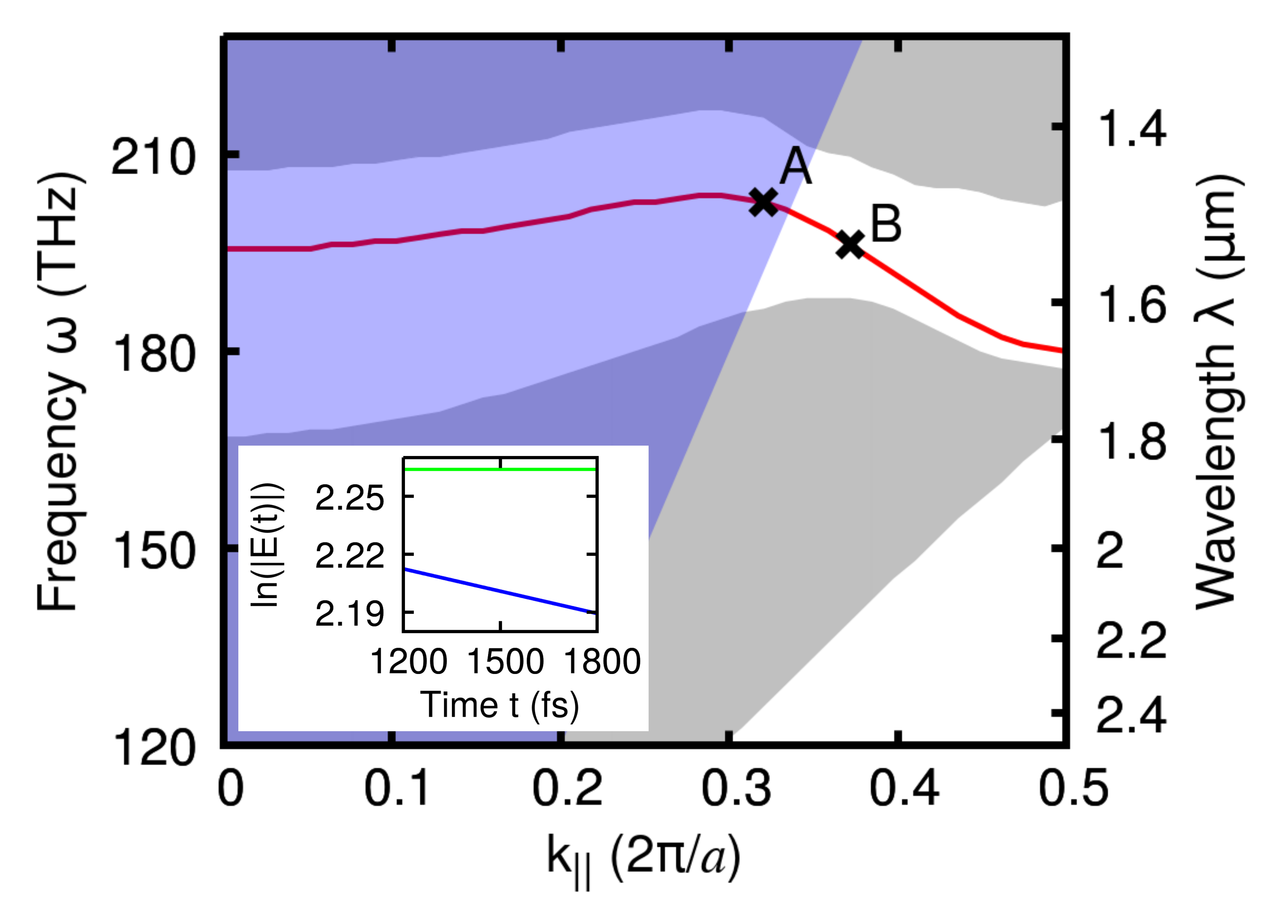}
\caption{
Calculated band structure of the PhC slab along the front, red, interface configuration (see Fig.~\ref{fig:supercell_band}(a)). 
The grey region is the projected bulk bandstructure, the red line is the edge mode dispersion and the purple shaded area represents the light cone. 
Modes at $A$ and $B$ are above and below the light line, respectively.
The inset shows the temporal decay of the mode at point $A$ (blue curve) and $B$ (green curve).
}
\label{fig:kagome_3D_band}
\end{figure}

\begin{figure}
\centering
\includegraphics[width=\columnwidth]{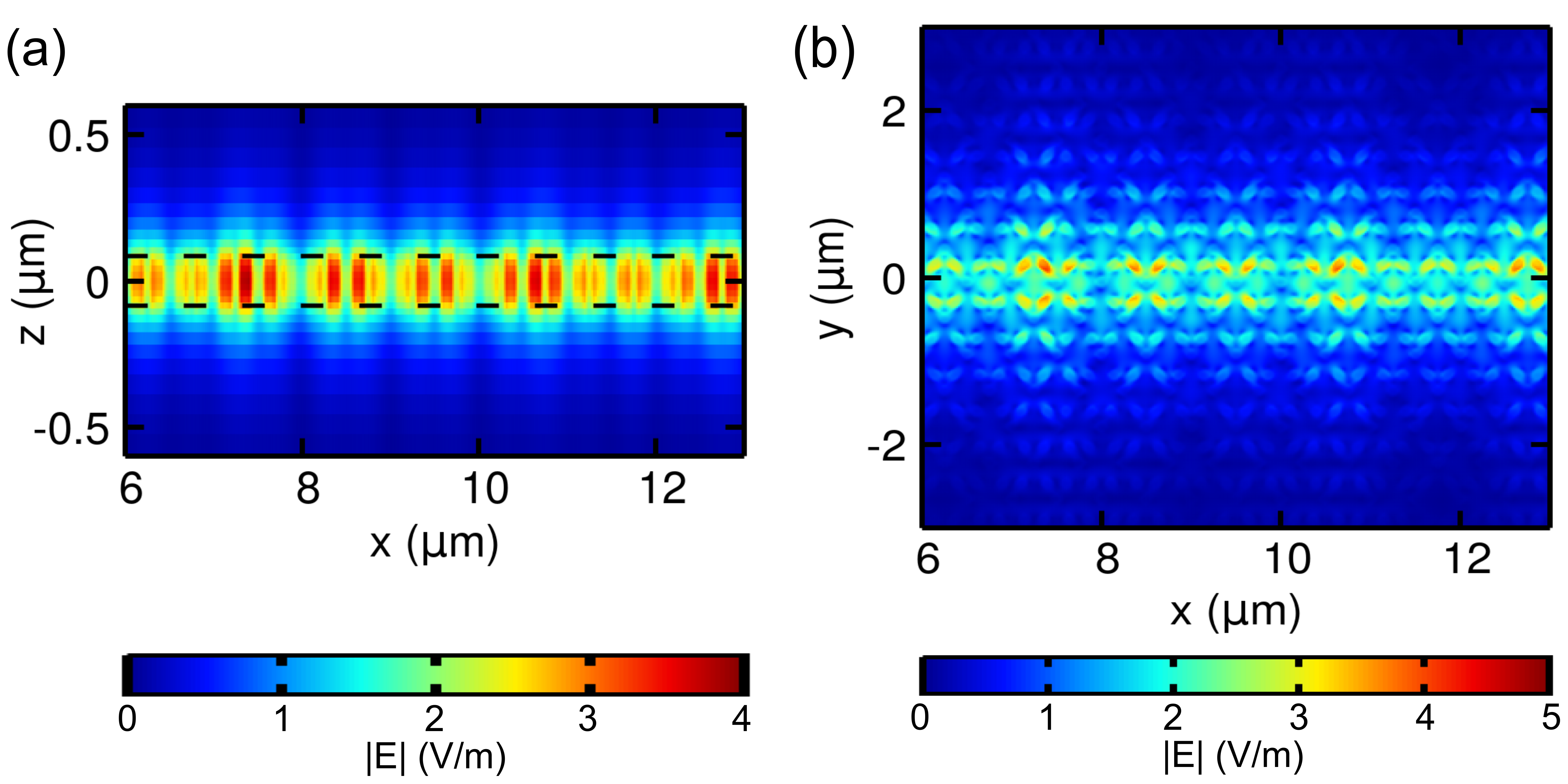}
\caption{
    Field profile of the edge modes lying below the light line (point $B$ in Fig.\ref{fig:kagome_3D_band}) in (a) the $y=0$ plane and (b) the $z=0$ plane.
The dashed lines are a guide to the eye for the $z$-limit of the free-standing slab.
}
\label{fig:kagome_3D_Efield}
\end{figure}

From a practical point of view, for the case of a $\Gamma {-} \text{K}$ interface inclination, edge modes lie close to or below the light line, thus improving vertical mode confinement without the need of sandwiching the waveguides with mirrors~\cite{Wu2015}. To make use of the better confinement, we propose a design composed of an InGaAsP free-standing PhC slab ($n = 3.3$~\cite{Kim2016a}) of $\SI{170}{\nano\metre}$ thickness. Fabrication of this structure can be carried out using a III${-}$V semiconductor wafer consisting of an InGaAsP substrate. The pattern of the air-holes can be defined by standard electron-beam lithography and inductively coupled plasma reactive-ion etching.
Importantly, the proposed structure can be designed to work at telecommunication wavelengths, from $\SI{1.2}{\micro\metre}$ to $\SI{1.8}{\micro\metre}$, as illustrated in Fig.~\ref{fig:kagome_3D_band}. The supercell band structure for TE-like modes (even-symmetric electric fields) is calculated with a finite-difference time-domain (FDTD) solver~\cite{Lumerical} 
for a lattice constant $a \eq \SI{0.5}{\micro\metre}$, air-holes with diameter $d \eq 0.3a \eq \SI{150}{\nano\metre}$ and perturbation $\delta \eq {\pm} \, 0.15$. The band gap ranges from $\SI{1.48}{\micro\metre}$ to $\SI{1.58}{\micro\metre}$ corresponding to a mid-gap at $\SI{1.53}{\micro\metre}$. 
Since the materials are assumed to be lossless in our simulations, any temporal decay, quantified by $\gamma$ in $E(t) \propto e^{i(\omega+i\gamma)t}$, is due to the leakiness of the mode. The inset of Fig.~\ref{fig:kagome_3D_band} shows the slope of the curves corresponding to $\gamma$, and demonstrating a negligible $\gamma_B$ for the mode below the light line (point $B$), whereas a finite leakage $\gamma_A$ is present for the mode lying above the light line (point $A$). 
Fig.~\ref{fig:kagome_3D_Efield}(a),(b) shows the electric field distribution corresponding to the point $B$ in Fig.~\ref{fig:kagome_3D_band}. Out-of-plane confinement is guaranteed by the mode's presence below the light line.

In summary, we have introduced a new kagome-like photonic topological insulator emulating the optical analogue of the quantum valley Hall effect. 
We have numerically shown that the associated topological edge modes do not intrinsically suffer from back-scattering for tailored inclinations. 
For $\Gamma {-} \text{K}$ inclination, the coupling between pseudo-spin channels is shown to be negligible while edge modes are guaranteed in the center of the band gap.
Based on our theoretical predictions, we have presented a realistic 3D design that can be fabricated with state-of-the-art methods~\cite{Kim2016a}, and works at suitable wavelengths for telecommunication applications. We have demonstrated improved vertical confinement due to edge modes lying below the light line.
The simplicity of the proposed design structure due to its monodisperse rods/holes makes it possible to fabricate it using many conventional techniques such as selective-area epitaxy or electron-beam patterning.
We have also shown that possible challenges resulting from low filling ratio can be overcome by tuning the perturbation making the band gap of about 100nm at telecommunication wavelengths.
%
Work has been done on kagome lattice in photonic systems~\cite{Ni2017_photonic}. Complementing previous work with kagome lattices, we have provided an in-depth analysis by explaining the reason of possible back-reflection translated in the transmission spectra and analysing the out-of-plane loss in 3D realistic design.



\begin{acknowledgments}
This work is part-funded by the European Regional Development Fund through the Welsh Government (80762-CU145 (East)).  We would like to thank Anthony J. Bennett, Yongkang Gong and Andreas Pusch for helpful discussions.
\end{acknowledgments}

\bibliographystyle{IEEEtran}
\bibliographystyle{apsrev4-1}
\bibliography{ref}

\appendix

\section{Numerical computation of the Berry curvature} 
\label{supp:berry_curv_comp}

The purpose of this appendix is to present the method used to calculate the Berry curvature.
Numerically, calculating the Chern number, or more generally the Berry curvature, is complicated if we do not have an easy closed-form analytical expression of the corresponding Hermitian operator. Indeed, for each $\vec{k}$ point, the eigenvectors carry a random phase $e^{i \phi(\vec{k})}\ket{\vec{u_k}}$ which is a numerical problem because of the derivative with respect to $\vec{k}$ appearing in the Berry curvature $\vec{\mathcal{F}(k)} \eq \nabla_{\vec{k}} \times i \braket{\vec{u_k}}{\nabla_{\vec{k}}\vec{u_k}}$.

\begin{figure}[t]
\center
\includegraphics[width=\columnwidth]{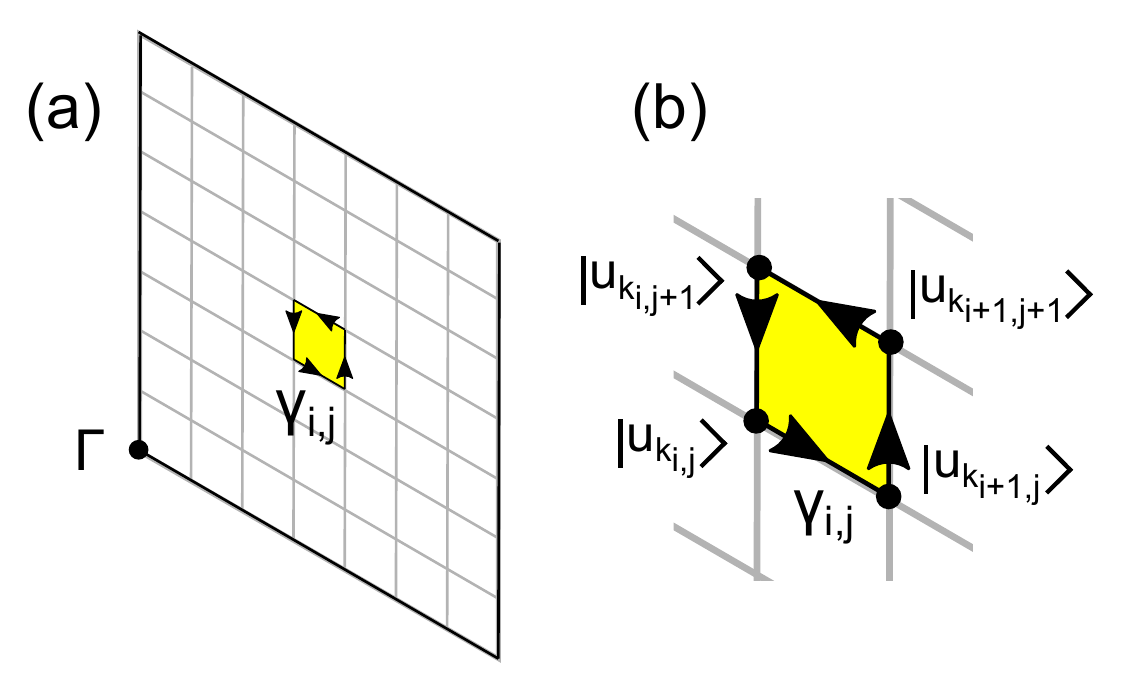}
\caption{
(a) Sketch of a closed path inside the (discretized) primitive unit cell in the reciprocal space. $\Gamma$ is the origin of the BZ: $\vec{k} \eq (0,0)$. 
$\gamma_{i,j}$ is the Berry phase defined along this loop and the yellow shaded area is the surface $S$ generated by this latter closed contour.
(b) Zoom in of the closed path. $\ket{\vec{u}_{\vec{k}_{i,j}}}$ are the eigenvectors at point $\vec{k}_{i,j}$ in the reciprocal space.
}
\label{fig:berry_phase_BZ}
\end{figure}

One solution is to start with the Berry phase $\gamma_{i,j}$ over a loop defined as in Fig.~\ref{fig:berry_phase_BZ}(a). Assuming the Berry connection is behaving well enough, one can apply Stokes' theorem which yields:
\begin{equation}
\gamma_{i,j} = \int_{S} \vec{\mathcal{F}(\vec{k})} \cdot d^2\vec{k}
= \mathcal{F}_{i,j} \; S
\end{equation}
where $S$ is the surface generated by the closed path and $\mathcal{F}_{i,j}$ is the Berry curvature defined at the $\vec{k}_{i,j}$ point (Fig.~\ref{fig:berry_phase_BZ}(b)) and assumed to be constant in the surface $S$.
Therefore, the Berry curvature can be seen as a Berry phase per unit area:
\begin{equation}
\mathcal{F}_{i,j} = \frac{\gamma_{i,j}}{S}
.
\label{eq:berry_curv}
\end{equation}

The problem is then reduced to calculating the Berry phase instead of the Berry curvature.
By definition, the Berry phase is the geometric phase that the eigenvector acquires after doing a loop in the parameter space. 
The phase difference $\Delta \phi_{\vec{k}}$ between two eigenvectors $\ket{\vec{u_k}}$ and $\ket{\vec{u_{k'}}}$ is: 
\begin{equation}
\Delta \phi_{\vec{k}} = \text{Im} \left[ \text{ln}(e^{i \Delta \phi_{\vec{k}}}) \right]
\end{equation}
with
\begin{equation}
e^{i \Delta \phi_{\vec{k}}} = \frac{\braket{\vec{u_k}}{\vec{u_{k'}}}}{ \Big| \braket{\vec{u_k}}{\vec{u_{k'}}} \Big| } 
.
\end{equation}
This means that, in this case (Fig.~\ref{fig:berry_phase_BZ}(b)):
\begin{multline}
\gamma_{i,j} = \text{Im} \Big[ \text{ln} \Big( \braket{\vec{u_{k_{i,j}}}}{\vec{u_{k_{i+1,j}}}} \braket{\vec{u_{k_{i+1,j}}}}{\vec{u_{k_{i+1,j+1}}}} 
\Big. \Big. \\ 
\Big. \Big.
\braket{\vec{u_{k_{i+1,j+1}}}}{\vec{u_{k_{i,j+1}}}} \braket{\vec{u_{k_{i,j+1}}}}{\vec{u_{k_{i,j}}}} \Big) \Big]
\label{eq:berry_phase}
\end{multline}
where the inner product is defined as~\cite{Joannopoulos2008}:
\begin{equation}
\braket{\vec{u_k}}{\vec{u_{k'}}} = \int \vec{u_k}^*(\vec{r}) \cdot \epsilon(\vec{r}) \vec{u_{k'}}(\vec{r}) \, d\vec{r}
\label{eq:scalar_prod_k}
\end{equation}
such that the photonic operator for the eigenvalue problem is Hermitian~\cite{Joannopoulos2008}.
One can note that the expression for the Berry curvature $\mathcal{F}_{i,j}$ (eqs.~\ref{eq:berry_curv}~and~\ref{eq:berry_phase}) is now gauge-independent, since the randomly k-dependent phases cancel each other in the $\ket{\vec{u_k}} \bra{\vec{u_k}}$ term. Moreover,  $\mathcal{F}_{i,j}$ converges to the continuous form when the spacing between neighbouring k-points approaches zero.

Once the Berry curvature $\mathcal{F}_{i,j}$ at the $\vec{k}_{i,j}$ point is obtained, one just needs to sum it around the desired surface to get the Chern number or the valley Chern number.

\section{Valley Chern number}
\label{supp:valley_chern_number}

In this section, we show that the valley Chern number $C_{\text{K}/\text{K}'}$ depends on the perturbation strength and explain why $C_{\text{K}/\text{K}'} \eq {\pm} \, 0.5$ is only achieved for infinitesimal perturbation.

The Weyl Hamiltonian in the vicinity of the K point takes the following form:
\begin{equation}
\mathcal{W} = \vec{h} \cdot \vec{\sigma}
\label{eq:Weyl_H}
\end{equation}
where $\vec{h} \eq (h_x, h_y, h_z) \eq (-\delta k_y, \tilde{\delta}, \delta k_x)$ is a vector in a 3D parameter space, $\delta k_i \eq k_i - k_{K, i} \, , \; i=x,y$ (Eq.~\ref{eq:H_kagome} in the main text).
This gives the following Berry curvature, for the lower band~\cite{Berry1984}:
\begin{equation}
\vec{\mathcal{F}(h)} = -\frac{1}{2} \frac{\vec{h}}{|\vec{h}|^3}
\label{eq:berry_curv_monopole}
\end{equation}
with $|\vec{h}| \eq \sqrt{\delta k_x^2 + \delta k_y^2 + \tilde{\delta}^2} \eq \sqrt{\vec{\delta k}^2 + \tilde{\delta}^2}$.
By integrating over a loop around K, $\gamma_\text{K}$, the valley Chern number is:
\begin{equation}
C_\text{K} = \frac{1}{2\pi} \int_{S_\text{K}} \vec{\mathcal{F}(h)} \, d^2\vec{k} = \frac{1}{2\pi} \int_{S_\text{K}} \frac{1}{2} \frac{\tilde{\delta}}{|\vec{h}|^3} \, dk_x \wedge dk_y
\end{equation}
where $d^2\vec{k} \eq dk_x \wedge dk_y$, $S_\text{K}$ is the surface generated by the closed path $\gamma_\text{K}$. This gives $C_\text{K} \eq {-} \, \text{sign}(\tilde{\delta}) \frac{1}{2}$. 

\begin{figure}
\center
\includegraphics[width=\columnwidth]{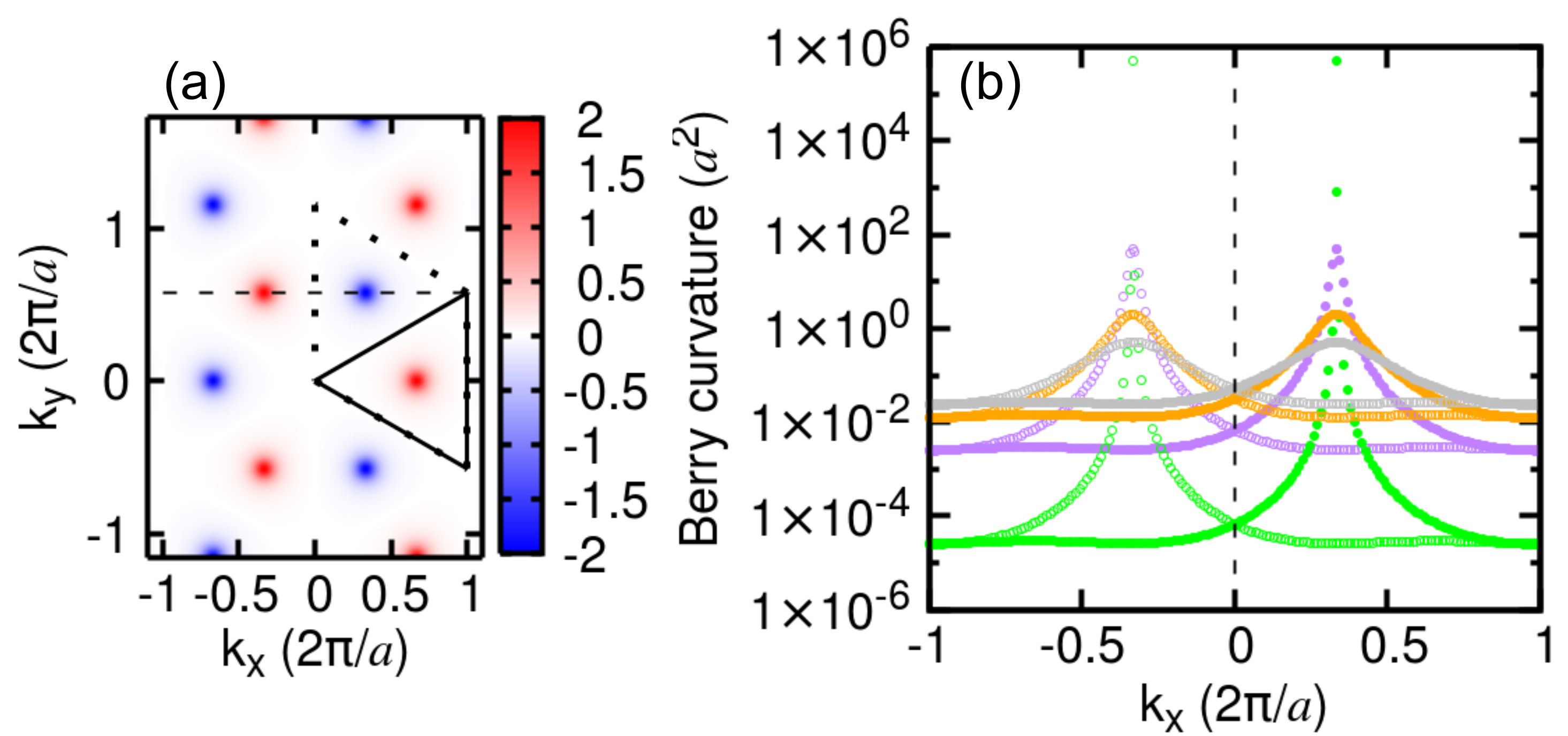}
\caption{
(a) Berry curvatures calculated using Eq.~\ref{eq:berry_curv_num} with perturbation $\tilde{\delta} \eq 0.5$. The dotted line shows the limit of one reciprocal primitive unit cell. The solid line represents the contour $\Gamma_K$ used to calculate the K-valley Chern number. The horizontal dashed line represents the path followed to plot (b).
(b) Plot of the absolute value of the Berry curvature for different perturbations: $\tilde{\delta} \eq 1e^{-3}$ (green circles), $\tilde{\delta} \eq 0.1$ (purple circles), $\tilde{\delta} \eq 0.5$ (orange circles) and  $\tilde{\delta} \eq 1$ (grey circles). The open and closed correspond to $\vec{\mathcal{\tilde{F}}}_\text{K}(\vec{\delta k}, \tilde{\delta})$ and $\vec{\mathcal{\tilde{F}}}_{\text{K}'}(\vec{\delta k}, \tilde{\delta})$ respectively. The vertical dashed line represents the boundary of the contour integration.
}
\label{fig:weyl_points}
\end{figure}

However, $C_\text{K} \eq {-} \, C_{\text{K}'}$ because of time-reversal symmetry, hence we need to take this opposite Weyl charge into account in the practical calculation: information from the positive and negative Weyl charges cannot be separated numerically.
Indeed, for numerical calculation, the loop is chosen, for simplicity, to be half the Brillouin zone, as depicted in Fig.~\ref{fig:weyl_points}(a). Therefore, when calculating the Berry curvatures, the Berry flux coming from the positive and negative charges cancel each other for high enough perturbation $\tilde{\delta}$.
Thus the valley Chern number becomes:
\begin{equation}
\tilde{C}_\text{K} =  \frac{1}{2\pi} \int_{S_\text{K}} \vec{\mathcal{\tilde{F}}(h)} \, d^2\vec{k}
\end{equation}
where, here, the tilde on $\tilde{C}_\text{K}$ and $\vec{\mathcal{\tilde{F}}(h)}$ stands for the practical value, numerically calculated, of the valley Chern number and Berry curvature, with:
\begin{equation}
\vec{\mathcal{\tilde{F}}}(\vec{\delta k}, \tilde{\delta}) = \vec{\mathcal{\tilde{F}}}_\text{K}(\vec{\delta k}, \tilde{\delta}) +\vec{\mathcal{\tilde{F}}}_{\text{K}'}(\vec{\delta k}, \tilde{\delta})
.
\label{eq:berry_curv_num}
\end{equation}
The positive and negative Weyl charges contribution at K and K$'$ are given by $\vec{\mathcal{\tilde{F}}}_\text{K}(\vec{\delta k}, \tilde{\delta})$ and $\vec{\mathcal{\tilde{F}}}_{\text{K}'}(\vec{\delta k}, \tilde{\delta})$ respectively:
\begin{equation}
\vec{\mathcal{\tilde{F}}}_\text{K}(\vec{\delta k}, \tilde{\delta}) =  \frac{\tilde{\delta}}{2} \sum_{(m, n) \in \mathrm{Z}^2} \left[ (\vec{\delta k}_{m, n})^2 + \tilde{\delta}^2 \right]^{-3/2} 
\label{eq:berry_curv_num_K}
\end{equation}
\begin{equation}
\vec{\mathcal{\tilde{F}}}_{\text{K}'}(\vec{\delta k}, \tilde{\delta}) = - \frac{\tilde{\delta}}{2} \sum_{(m, n) \in \mathrm{Z}^2} \left[ (\vec{\delta k}_{1/2 + m, 1/2 + n})^2 + \tilde{\delta}^2 \right]^{-3/2} 
\label{eq:berry_curv_num_Kprime}
\end{equation}
where $\vec{\delta k}_{m,n} \eq \vec{k} \, {+} \, m \vec{b_1} \, {+} \, n \vec{b_2} \, {-} \, \vec{k_\text{K}}$, with the reciprocal lattice vectors $\vec{b_i}$.

Fig.~\ref{fig:weyl_points}(a) shows the calculated Berry curvature using Eq.~\ref{eq:berry_curv_num} with perturbation $\tilde{\delta} \eq 0.5$. The dotted lines correspond to one reciprocal primitive unit cell and the solid lines represent the contour integral $\Gamma_\text{K}$ performed to calculate the K-valley Chern number. The horizontal dashed line represents the path followed to plot the absolute of the Berry curvatures $\vec{\mathcal{\tilde{F}}}_\text{K}(\vec{\delta k}, \tilde{\delta})$ and $\vec{\mathcal{\tilde{F}}}_{\text{K}'}(\vec{\delta k}, \tilde{\delta})$ in Fig.~\ref{fig:weyl_points} in open and closed circles, respectively.
In Fig.~\ref{fig:weyl_points}(b), the Berry curvature is plotted for $\tilde{\delta} \eq 1e^{-3}$ (green circles), $\tilde{\delta} \eq 0.1$ (purple circles), $\tilde{\delta} \eq 0.5$ (orange circles) and  $\tilde{\delta} \eq 1$ (grey circles). The vertical dashed line represents the boundary of the contour integration.
This illustrates that for infinitesimal small perturbation, the ``leakage" of the Berry flux can be negligible compared to its high value at the K/K$'$ points and one get $\tilde{C}_K \eq 0.5$. In contrast, when the perturbation is relatively high, \ie not infinitesimal, the Berry flux associated to Weyl charges are leaking out of the more or less arbitrarily defined valley domain $\Gamma_{\text{K}/\text{K}'}$ respectively, and are cancelling with the Berry flux of opposite Weyl charges. For $\Gamma_{\text{K}/\text{K}'}$ defined as in Fig.~\ref{fig:weyl_points}, this results in lower valley Chern number than expected: $\tilde{C}_K \eq 0.47$ for $\tilde{\delta} \eq 0.1$, $\tilde{C}_K \eq 0.36$ for $\tilde{\delta} \eq 0.5$ and $\tilde{C}_K \eq 0.25$ for $\tilde{\delta} \eq 1$.
This shows that the valley Chern number is not a proper topological invariant since the sign of the perturbations is the same, \ie the gap does not close, but the valley Chern number calculated changes for different perturbations.

The angular distribution of the Berry curvature obtained here in this section is obviously different from the one obtained with MPB. 
The difference is rooted in the fact that the Weyl Hamiltonian (Eq.~\ref{eq:Weyl_H}) is only a first order approximation of perturbations in any direction, i.e. not valid for finite geometrical perturbation. Of course that does not change the valley Chern number for any surface enclosing only one Weyl monopole, \ie the integrated Berry flux stays the same, but the angular distribution of the Berry flux will generally be altered (see figure of the Berry curvature in the main text). The cancellation of Berry flux associated with opposite Weyl charges will therefore be different.
However, as stressed in the main text, only the sign matters in determining the topology and possible topological edge modes.

\section{Transmission}
\label{supp:transmission}

The purpose of this section is to explain the presence of the oscillations in the transmission spectra obtained in the main text.

\begin{figure}[t]
\center
\includegraphics[width=\columnwidth]{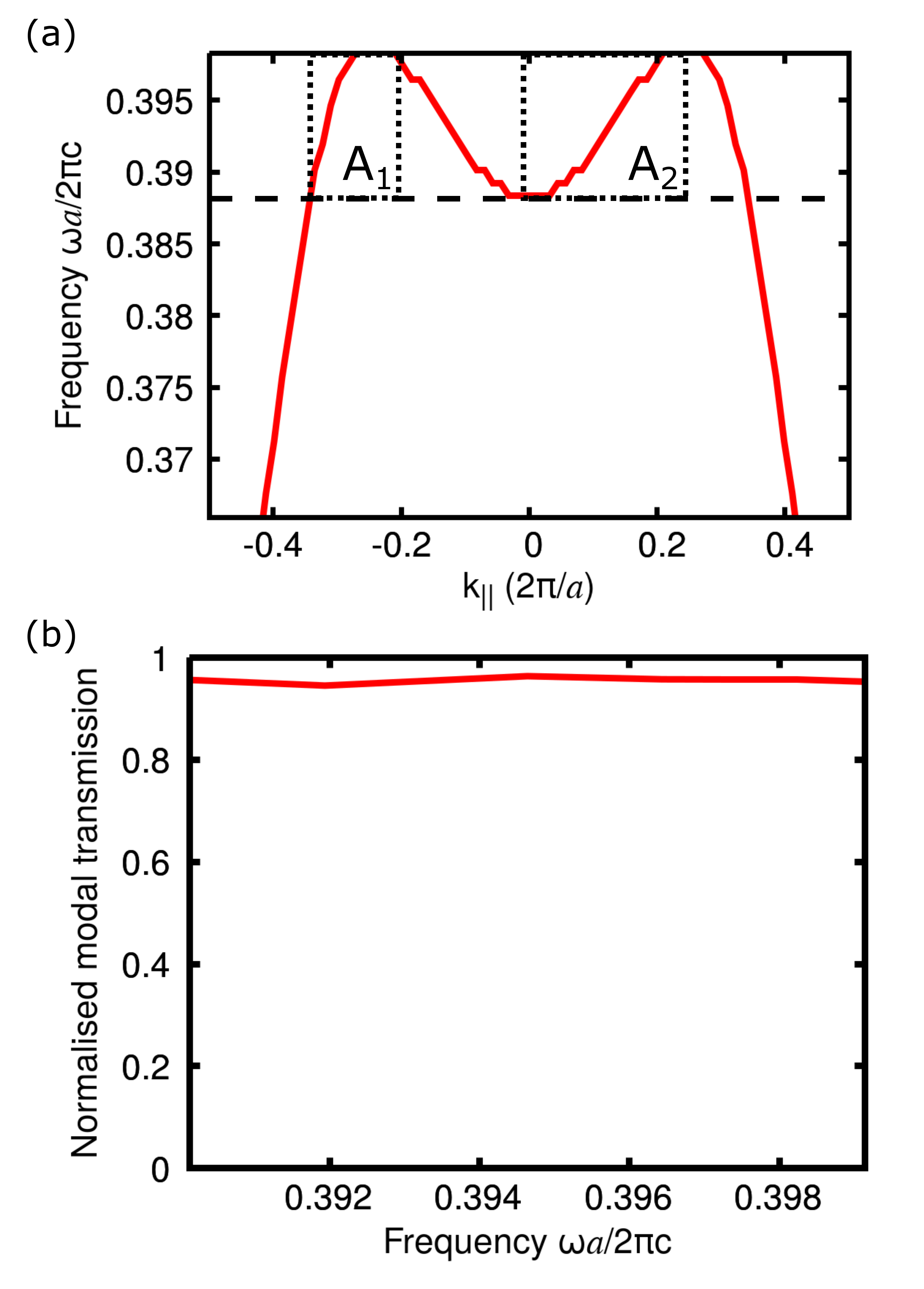}
\caption{
(a) Dispersion of the edge modes at the red interface (see main text) within the band gap. The dashed lines correspond to the frequency range above which several modes have the directionality. The dotted boxes correspond to the modes ($A_1$ and $A_2$) propagating to the right.
(b) Normalised modal transmission of the mode $A_1$ within the frequency range of multimode regime in (a).
}
\label{fig:supp_modal_transmission}
\end{figure}

Oscillations observed in the transmission spectra of the waveguides can be explained because of the coupling of the Gaussian source with the waveguided modes.
Figure~\ref{fig:supp_modal_transmission}(a) shows the FDTD-calculated dispersion of the edge modes along the red interface (see main text). There exists a frequency range where several mode can be excited for a given frequency. Looking at the modes propagating on the right, one can excite two edge modes $A_1$ and $A_2$ with positive group velocity. The electric $E(x, \omega)$ field therefore has the following form:
\begin{equation}
E(x, \omega) = A_1(\omega) e^{i k_1 x} u_{k_1}(x) + A_2(\omega) e^{i k_2 x} u_{k_2}(x).
\end{equation}
The transmission $T$ is given by:
\begin{equation}
T(x, \omega) = \left| \frac{E_t(x, \omega)}{E_0(x, \omega)} \right|^2
\end{equation}
where $E_t(x, \omega)$ and $E_0(x, \omega)$ are the transmitted and incident electric field, respectively. 
Therefore, the oscillations are coming from the individual modal transmissions coefficient $t_i(\omega)$ of the mode $A_i, \, i=1,2$:
\begin{equation}
t_i(\omega) = \frac{A_{i,t}(\omega)}{A_{i,0}(\omega)}
\end{equation}
where the $A_{i,t}(\omega)$ and $A_{0,t}(\omega)$ stand for the transmitted and incident modal amplitudes, respectively.
The normalised modal transmission $T_i = |t_i|^2$ is plotted in Fig.~\ref{fig:supp_modal_transmission} for the mode $A_1$ within the multimode frequency regime and the waveguide shown in the main text (Z-shaped waveguide along the red interface).
Looking at the modal transmission, this shows clearly the high transmission for this mode close to the $K$ (or $K'$) point.

\section{Chirality map}
\label{supp:chirality_map}

\begin{figure}[t]
\center
\includegraphics[width=\columnwidth]{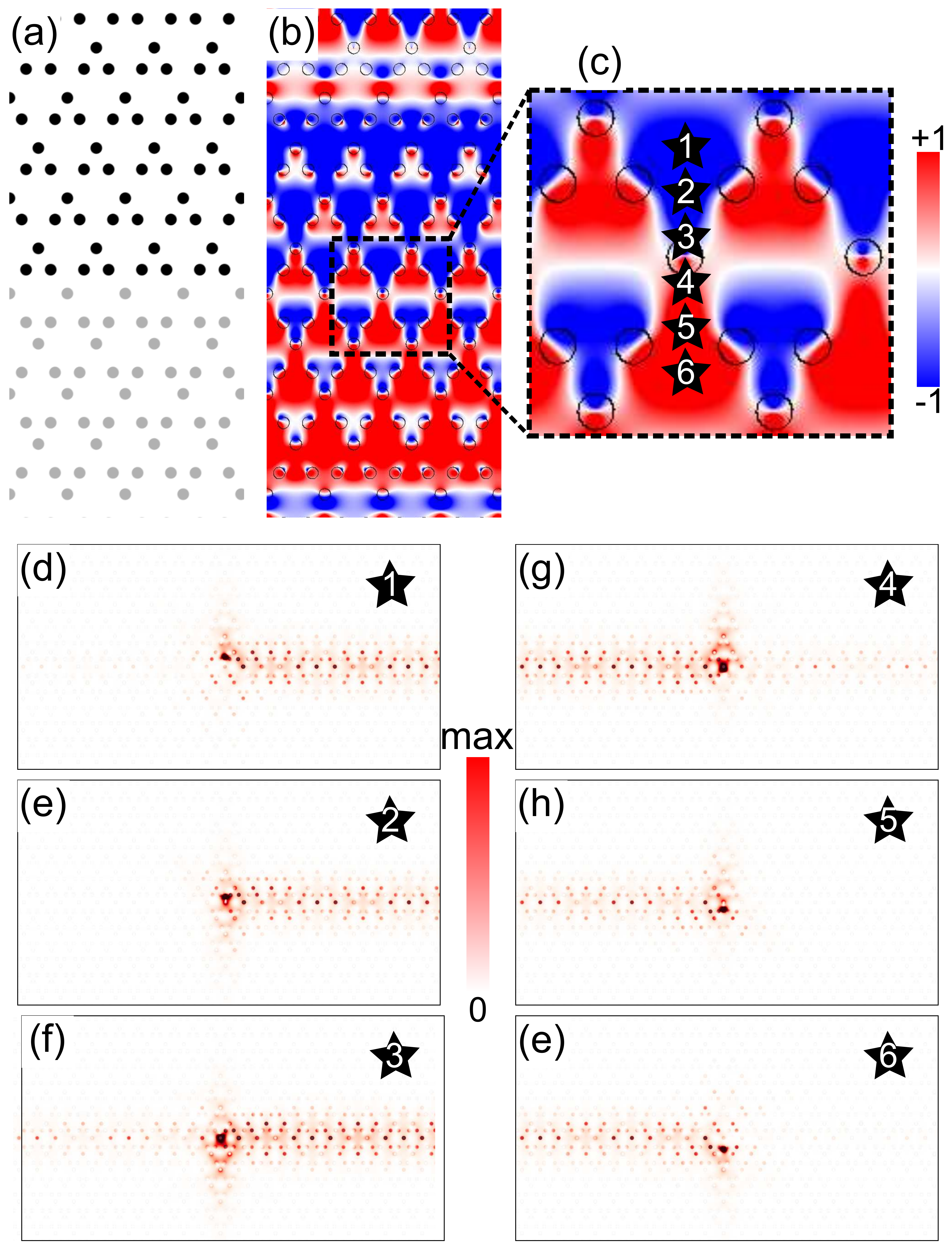}
\caption{
(a) Sketch of the interface considered between two opposite kagome lattice.
(b) Chirality map calculated at the $k_\parallel \eq 0.4$ point.
(c) Zoom in of (c) with different positions of the source considered represented by the star.
(d)-(e) Total power plot of the edge mode for the different positions of the source with polarized source $S_+ \eq H_x + i H_y$.
}
\label{fig:chirality_transition}
\end{figure}

The statement made in the main text on unidirectional pseudo-spin excitation is demonstrated by the full wave dynamics of the system. The source chosen is left-polarized $S_+ \eq H_x + i H_y$ and only one pseudo-spin unidirectional edge mode can be excited by locating the source at a certain position. The information is given by means of a wave-vector $k_\parallel$-dependent chirality map~\cite{Oh2018}. Fig.~\ref{fig:chirality_transition}(a) shows the interface considered: it corresponds to the red interface in (see main text) with perturbation $\delta \eq {\pm} \, 0.15$. Therefore we will look at only the edge modes represented by the red dispersion line.

Transverse magnetic (TM) modes (non-zero $E_z, H_x, H_y$) are considered here but the concept of chirality of the edge modes can be similarly applied to  transverse electric (TE) modes (non-zero $H_z, E_x, E_y$).
From the magnetic field components $(H_x, H_y)$ of the edge modes, the chirality of the mode for H field can be calculated using the Stokes parameters. Fig.~\ref{fig:chirality_transition}(b) shows the calculated chirality at the corresponding $k_\parallel \eq 0.4$ point. It is interesting to see that the chirality map has a region between the two different PTIs with almost the same sign of values (close to $+1$ ($-1$), shown in red (blue) in Fig.~\ref{fig:chirality_transition}(b)). This implies that different circular polarizations are needed to excite edge modes propagating in the same direction at the two positions (\eg points denoted by 1 and 6 in Fig.~\ref{fig:chirality_transition}(c)). Alternatively, the same circularly polarized dipoles at the two positions would excite edge modes propagating in the opposite direction. 
This is shown in Fig.~\ref{fig:chirality_transition}(d)-(e) where the total power is plotted for different source positions 1 to 6 (Fig.~\ref{fig:chirality_transition}(c)), with polarization $S_+ \eq H_x + i H_y$. Starting with the source located at the position 1, \ie in the negative chirality, the excited topological edge mode corresponds to the one propagating to the right. Moving the location of the source to position 6 with positive chirality will predominantly excite the mode propagating to the left. Therefore the position of the source is crucial for the excitation of unidirectional topological edge modes, as in other systems~\cite{Petersen2014, Coles2016, Lodahl2017, Barik2018} with the main difference being the robustness to bendings.

\end{document}